\documentclass[conference]{IEEEtran}
\IEEEoverridecommandlockouts

\usepackage{cite}
\usepackage{amsmath,amssymb,amsfonts}
\usepackage{algorithm, algorithmic}
\usepackage{graphicx}
\usepackage{textcomp}
\usepackage{xcolor}
\usepackage{bm}
\usepackage{url}
\usepackage{adjustbox}

\usepackage[hidelinks]{hyperref}

\usepackage{subcaption}

\usepackage{enumitem}

\newboolean{showcomments}
\setboolean{showcomments}{false} 

\ifthenelse{\boolean{showcomments}}
  {
    \newcommand{\nb}[2]{
      \fcolorbox{gray}{yellow}{\bfseries\sffamily\scriptsize#1}
      {$\blacktriangleright$#2$\blacktriangleleft$}
    }
    \newcommand{\oshani}[1]{{\color{red}\bfseries [Oshani: #1]}}
    \newcommand{\saikat}[1]{{\color{green}\bfseries [saikat: #1]}}
    \newcommand{\sep}[1]{{\color{blue}\bfseries [sep: #1]}}
    
  }
  {
    \newcommand{\nb}[2]{}
    \newcommand{\oshani}[1]{}
    \newcommand{\saikat}[1]{}
    \newcommand{\sep}[1]{}
    
  }

\newtheorem{theorem}{Theorem}[section]

\newtheorem{definition}[theorem]{Definition}

\makeatletter
\def\@begintheorem#1#2{\trivlist
   \item[\hskip \labelsep{\bfseries #1\ #2.}]\itshape}
\makeatother

\DeclareMathOperator{\Db}{\textbf{D}}
\DeclareMathOperator{\Xb}{\textbf{X}}
\DeclareMathOperator{\xb}{\textbf{x}}
\DeclareMathOperator{\eb}{\textbf{e}}
\DeclareMathOperator{\yb}{\textbf{y}}
\DeclareMathOperator{\thetab}{\bm\theta}
\DeclareMathOperator{\Thetab}{\bm\Theta}
\DeclareMathOperator{\Loss}{\mathcal L}
\DeclareMathOperator{\B}{\mathcal B}

\newcommand{\eqdef}{\ensuremath{\stackrel{\text{def}}{=}}}
\newcommand{\expec}{\ensuremath{\mathbb E}}

\usepackage{cleveref}

\usepackage{flushend}

\begin{document}

\makeatletter
    \newcommand{\linebreakand}{%
      \end{@IEEEauthorhalign}
      \hfill\mbox{}\par
      \mbox{}\hfill\begin{@IEEEauthorhalign}
    }
\makeatother

\title{Fed-RD: Privacy-Preserving Federated Learning for Financial Crime Detection%
\thanks{The authors acknowledge the support from NSF IUCRC CRAFT center research grant (CRAFT Grant \# 22009) for this research. The opinions expressed in this publication do not necessarily represent the views of NSF IUCRC CRAFT.
We are also grateful for the advice from our CRAFT Industry Advisory Board members in shaping this work, especially the input from Richard Hoehne from IBM, and Chalapathy Neti and Ravi Doddasomayajula from SWIFT.
}
}

\author{\IEEEauthorblockN{Md. Saikat Islam Khan\IEEEauthorrefmark{1}, Aparna Gupta\IEEEauthorrefmark{2}, Oshani Seneviratne\IEEEauthorrefmark{1}, and Stacy Patterson\IEEEauthorrefmark{1}}%
\IEEEauthorblockA{\IEEEauthorrefmark{1}Department of Computer Science, \IEEEauthorrefmark{2}Lally School of Management\\
Rensselaer Polytechnic Institute, Troy, NY, USA \\
Email: islamm9@rpi.edu, guptaa@rpi.edu, senevo@rpi.edu, sep@cs.rpi.edu}}

\maketitle

\begin{abstract}
We introduce Federated Learning for Relational Data (Fed-RD), a novel privacy-preserving federated learning algorithm specifically developed for financial transaction datasets partitioned vertically and horizontally across parties. Fed-RD strategically employs differential privacy and secure multiparty computation to guarantee the privacy of training data. We provide theoretical analysis of the end-to-end privacy of the training algorithm and present experimental results on realistic synthetic datasets. Our results demonstrate that Fed-RD achieves high model accuracy with minimal degradation as privacy increases, while consistently surpassing benchmark results.
\end{abstract}


\section{Introduction}
Financial crime poses an increasing threat to global safety and security, both physical and financial. According to UN estimates~\cite{4}, between 80 and 200 billion USD is laundered every year, with these illicit funds being used to facilitate crimes such as illegal arms sales, human trafficking, and terrorism. Further, a recent Nasdaq Verafin report~\cite{nasdaq2024} estimates  485.6 billion USD in total losses in 2023 from fraud scams and bank fraud schemes. This immense impact necessitates new technologies that can efficiently and accurately detect and help prevent criminal financial activity.

The vast amounts of data collected on financial customers and transactions provide a rich source of information to construct machine learning (ML) models that can detect anomalous financial transactions connected to criminal activity. Although various machine-learning techniques for financial fraud detection have been explored~\cite{alsuwailem2023performance,yi2023fraud,fanai2023novel}, they rely on centralized methods where all data management and training processes are consolidated into a central hub.
This centralization poses serious threats to the confidentiality of participant data and may lead to potential privacy breaches \cite{rahman2023federated}.
Such vulnerabilities can have particularly adverse effects in domains like finance, where maintaining the privacy of various parties is imperative, and financial organizations might be reluctant to share their data due to regulatory constraints and concerns about losing competitive advantage~\cite{van2023markets}. These conflicting demands present a challenge in striking a balance between facilitating adequate utilization of ML methodologies for detecting anomalous financial behavior and restricting the leakage of confidential data that could occur through data sharing.

A promising option is to employ federated learning (FL), a distributed machine learning paradigm where parties jointly train a global model without directly sharing raw data~\cite{mcmahan2017communication}. In FL algorithms, parties instead exchange intermediate updates, such as model gradients, throughout the training process. However, restricting access to the raw data is not sufficient to protect privacy, and it has been shown that information can be leaked through the intermediate updates~\cite{luo2021feature, yazdinejad2023ap2fl, sanchez2024federatedtrust}.
Such leakage could reveal details about individual transactions, account holders, or financial patterns, leading to privacy breaches and regulatory non-compliance. 
Furthermore, the majority of FL methods assume that the training data is partitioned in one of two ways: horizontally, where the parties share the same feature space \cite{mcmahan2017communication,zhang2023multi}, or vertically, where each party has a distinct feature space for the same set of sample IDs \cite{castiglia2023flexible,liu2024vertical}. These assumptions may not hold in complex real-world scenarios, such as financial crime detection. Thus, we need a privacy-preserving FL approach that is targeted specifically for financial transaction data.

We propose Federated Learning for Relational Data (Fed-RD), a privacy-preserving model training algorithm for detecting anomalous
 financial transactions. 
Fed-RD addresses training data that is partitioned between a \emph{transaction party} that stores features about individual financial transactions and a set of \emph{account parties}, e.g., banks, that hold information about the accounts involved in these transactions. This data arrangement is novel to FL in that it includes a many-to-many vertical partitioning between the transaction dataset and the account dataset, with one transaction corresponding to a sender and receiver account and one account corresponding to one or more transactions. It also includes horizontal partitioning of the account dataset across banks. To safeguard the privacy of participant data, Fed-RD makes targeted use of differential privacy (DP)~\cite{dwork2014algorithmic} and secure multiparty computation (MPC)~\cite{cramer2015secure} at information leak points in the algorithm. Our approach provides provable privacy protection with minimal degradation of accuracy in the final model. 

The key contributions of this work are as follows:
\begin{itemize}
\item We propose Fed-RD, which enables distributed model training over financial transaction data that is vertically and horizontally partitioned across parties.
\item We give formal definitions of data privacy targeted for FL over distributed financial transaction data.
 \item We provide formal guarantees of end-to-end data privacy for Fed-RD.
 \item We demonstrate the effectiveness of Fed-RD in experiments with realistic synthetic datasets. Our results show that Fed-RD achieves strong performance, with minimal accuracy loss as privacy parameters increase, while surpassing benchmark results.
\end{itemize}

\noindent \textbf{Paper outline:} The rest of the paper is organized as follows: Section~\ref{sec:background} provides a brief overview of DP and the privacy mechanisms Fed-RD employs.  We present the system model, training problem, threat model, and privacy objectives in Section~\ref{sec:Problem Formulation}. Section~\ref{sec:Proposed Algorithm} details our proposed algorithm, and Section~\ref{privcy} gives the formal privacy analysis. Experimental results are presented in Section~\ref{sec:Experiments}. Section~\ref{relatedwork} summarizes related work, and finally, we conclude in Section~\ref{sec:Conclusion} with the future outlook of this work.

\section{Background}
\label{sec:background}
In this section, we present background on DP, along with the pertinent mechanisms incorporated into the FED-RD algorithm to maintain privacy.

\subsection{Differential Privacy}
DP is an established method that provides theoretical privacy guarantees for datasets. 
\emph{Standard DP}~\cite{dwork2014algorithmic} is a randomized mechanism used to generate the output of a computation, for example, computing a sum, in a manner that obfuscates whether a particular input was used to produce the computation output. \emph{Local DP}~\cite{alvim2018local} is a variant of DP in which the data itself is obscured via a randomized process, providing privacy guarantees before the data is used in the computation. 

In Fed-RD, we utilize  \emph{R\'{e}nyi Differential Privacy (RDP)}~\cite{mironov2017renyi}, a specialized version of DP derived from the principle of  R\'{e}nyi divergence, to provide standard DP and local DP at various steps in our algorithm. 
We provide the formal definitions of R\'{e}nyi divergence and RDP below.
\begin{definition}[{R\'{e}nyi divergence}] For $\alpha \in (0, \infty), {\alpha \neq 1}$, the \textit{R\'{e}nyi divergence} of order $\alpha$ between two probability distributions ${\cal P}$ and ${\cal Q}$ is
\[
D_{\alpha}(P \| Q) \eqdef \frac{1}{\alpha - 1} \log \left ( \expec_{x \sim {\cal Q}}\left[{\cal P}(x)^{\alpha} {\cal Q}(x)^{1-\alpha}\right] \right)
\]
where $\log(\cdot)$ is the natural logarithm. 
\end{definition}
\begin{definition}[R\'{e}nyi Differential Privacy (RDP)] A randomized mechanism ${\cal M}$ satisfies ($\alpha, \epsilon$)-\emph{RDP} if for any two adjacent datasets $D$ and $D'$ and any $\alpha > 1$, it holds that
\[
D_{\alpha}(M(D) \| M(D')) \leq \epsilon.
\]
\end{definition}
Here, \emph{adjacent} means that datasets differ in exactly one record. 
The value of $\epsilon$, commonly called the \emph{privacy budget}, indicates the level of privacy for the dataset, with a smaller $\epsilon$ corresponding to greater privacy.
Informally, RDP provides a degree of indistinguishability of whether a particular record was included in the computation of ${\cal M}$.
\begin{definition}[Local R\'{e}nyi Differential Privacy (Local RDP)] A randomized mechanism ${\cal M}$ satisfies ($\alpha, \epsilon$)-\emph{Local RDP} if for any two inputs $x_1, x_2 \in {\cal X}$ it holds that
\[
D_{\alpha}({\cal M}(x_1) \| {\cal M}(x_2)) \leq \epsilon.
\]
\end{definition}
In Local RDP, $\epsilon$ quantifies the privacy that a mechanism ${\cal M}$ provides when applied to a single record.

We employ RDP because it facilitates the computation of cumulative privacy loss over the composition of multiple randomized mechanisms. Through this composition, we can provide theoretical bounds on the total privacy loss over the entire training process. 

\subsection{Gaussian Mechanism}  
In Fed-RD, we provide Local RDP  using a \emph{Gaussian mechanism}. To protect a scalar real-valued input $x$, 
Gaussian noise is added to produce the output as ${\cal G}(x) = x + \mathcal{N}(0, \sigma^2)$.
For $x \in [-1, 1]$, it has been shown that this mechanism provides $(\alpha, \epsilon(\alpha)$)-Local RDP with  $\epsilon(\alpha) = \frac{\alpha}{2\sigma^2}$~\cite{mironov2017renyi}.

\begin{algorithm}
    \caption{Poisson Binomial Mechanism} \label{pbm.alg}
    \label{alg:scalar_poisson_binomial}
    \begin{algorithmic}[1]
    \STATE {\textbf{Initialize:}} $x_i \in [-k, k]$; $b \in \mathbb{N}$, $\beta \in [0, \frac{1}{4}]$.
    \STATE Determine probability $p_i \leftarrow \frac{1}{2} + \frac{\beta}{k}x_i$.
    \STATE Generate quantized value $q_i \leftarrow \text{Binom}(b, p_i)$.
    \STATE \textbf{Output:} Quantized value $q_i$.
    \end{algorithmic}
\end{algorithm}
\subsection{Poisson Binomial Mechanism (PBM)} \label{sec:PBM}

Fed-RD uses the Poisson Binomial Mechanism (PBM) proposed in~\cite{chen2022poisson1} to provide standard DP over sum and average computations.
We briefly describe the process for computing the sum over a set of $M$ scalar values ${x_1, \ldots x_M}$, with each $x_i \in [-C, C]$, where each value is held by a different party. 
Each party first quantizes its value into one of $b$ bins according to Algorithm~\ref{pbm.alg}. 
The parties use MPC to find the sum $\hat{q} = \sum_{i=1}^M q_i$, while keeping each $q_i$ secret.
An estimated value of the sum is then computed from $\hat{q}$ as ${\tilde{s} =  \frac{C}{\beta b} (\hat{q} - \frac{bM}{2})}$.

Note that the PBM injects randomization by drawing values from a binomial distribution; thus, PBM provides quantization and randomization. The parameters $b$ and $\beta$ determine the degree of privacy.
It has been shown that the sum computation yields an unbiased estimate of the correct sum with variance $\frac{C^2M}{4\beta^2 b}$ while providing $(\alpha, \epsilon(\alpha))$-RDP  with $\epsilon(\alpha)~=~\Omega (b\beta^2 \alpha / (M-1))$~\cite{chen2022poisson1, tran2024differentially}.

\subsection{Multiparty Computation (MPC)}
\label{sec:MPC}

The privacy guarantees for standard RDP, such as the sum computation,  apply when only the sum is revealed. If any of the inputs to the sum are also shared, the privacy decreases. Thus we must ensure they remain private during the sum computation. We achieve this using MPC.

MPC is a cryptographic protocol that allows multiple parties to collaboratively compute a function over their inputs while ensuring that these inputs remain confidential.
We refer the reader to~\cite{bonawitz2016practical} for MPC protocols that can be applied in FL. One key requirement of MPC protocols is that the inputs must be integer values. Thus, PBM is a natural fit with MPC since it quantizes the inputs before the sum computation.

\begin{figure}
\centering
\begin{adjustbox}{max width=\textwidth}
\includegraphics[scale=0.16]{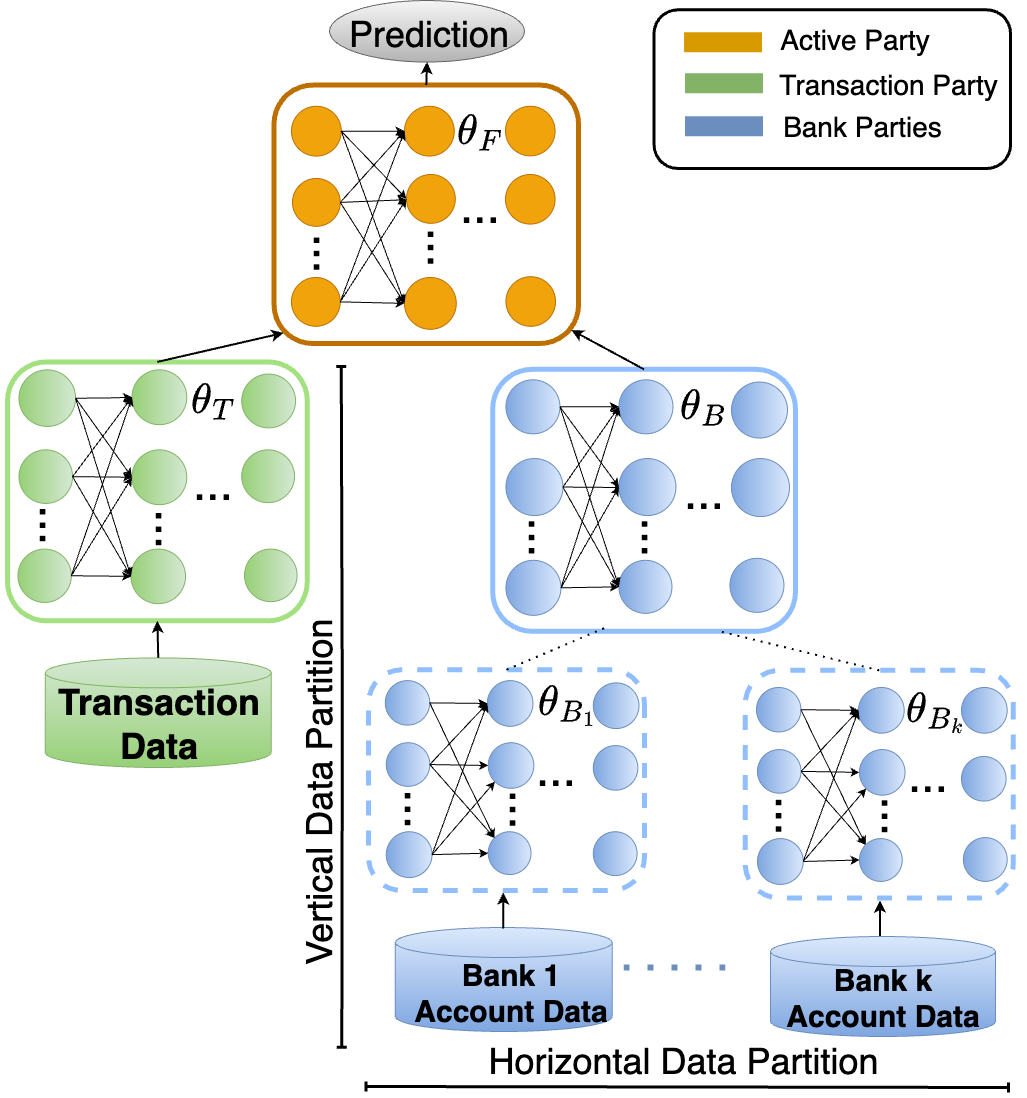}
\end{adjustbox}
\caption{Illustration of data and model partitioning among the transaction party and banks.}
\label{data.fig}
\vspace{-3ex}
\end{figure}

\section{Problem Formulation}
\label{sec:Problem Formulation}
\subsection{System Model}
We consider a distributed computing system comprised of two data silos. The first silo, the transaction silo $T$, has a single party that stores information about financial transactions, such as wire transfers between bank accounts. Its features include country, currency, ordering account, beneficiary account, etc.

The second silo, the bank silo $B$, maintains data about the accounts involved in these transactions. Silo $B$ comprises $K > 1$ parties. Each party (bank) stores information for a disjoint subset of accounts, such as account identifiers, flags, and addresses.
A one-to-many relationship exists between transactions and accounts, each transaction linking to two accounts, a sender and a receiver. Additionally, there is a one-to-many relationship between accounts and transactions; a single account may participate in multiple transactions, either as a sender or a receiver.
Our system also contains an \emph{active party} that holds the labels for the training data; a transaction is labeled 1 if it is anomalous and 0 otherwise. If the transaction party holds the labels, it plays the role of the active party.

\subsection{Training Problem}
We seek to develop an FL solution that enables the parties to collaboratively train a model to classify anomalous transactions. 
We let $\Xb_T$ denote the dataset consisting of the $N$ transactions held by silo $T$, and we let $\Xb_B$ denote the dataset of accounts held by silo $B$. Each transaction sample $\xb \in \Xb_T$ corresponds to two samples in $\Xb_B$, a sender account $\xb_s$ and a receiver account $\xb_r$. These account samples are held by different banks in silo $B$.
Each transaction sample $\xb \in \Xb_T$ also has a label $y$. Let $\yb$ denote the set of all labels.

The goal is to train an ML model over the training data $\Xb_B$, $\Xb_T$, and labels $\yb$ while protecting the privacy of the training data. The model consists of three components: a \emph{transaction model} that accepts a transaction sample as input and produces a \emph{transaction embedding}; an \emph{account model} that accepts an account sample as input and produces an \emph{account embedding}, and a \emph{fusion model} that accepts a transaction embedding, a sender account embedding, and a receiver account embedding as input and produces a prediction. We propose two variations for the fusion model; in the first, the fusion model accepts a concatenation of the three embeddings, and in the second, it accepts a sum of the three embeddings. We discuss these two approaches and their tradeoffs in the following sections.
We place no restrictions on the types of models; they may be simple, such as linear models, or more complex, such as neural networks.

The transaction party in $T$ is responsible for training the transaction model parameters $\thetab_T$. The bank parties in  $B$ are responsible for training the account model $\thetab_B$.
Each bank party holds a copy of the account model and collaborates to update its parameters. 
The fusion model $\thetab_F$ is trained by the active party. 
An illustration of the data and model architecture is shown in Figure~\ref{data.fig}.

Let $\Thetab$ denote the set of all model parameters.
Let $h_B(\thetab_B, \xb_b)$ represent the account model as a function that produces an embedding from the account $\xb_b$,  and let $h_T(\thetab_T, \xb_t)$ represent the transaction model as a function that produces an embedding from the transaction $\xb_t$. We assume each embedding is of length $P$.
To train the entire model, the parties collaborate to minimize a loss function of the form
\begin{align}
&\Loss(\Thetab; \Xb_T, \Xb_B, \yb) = \nonumber \\
&\sum_{\xb_t \in \Xb_T} \ell\left(\thetab_F, h_T(\theta_T; \xb_t), h_B(\theta_B, \xb_s), h_B(\theta_B; \xb_r); y_t \right)
\end{align}
where $\ell(\cdot)$ is the loss for a single transaction 
$(\xb_t, y_t)$ with  sender account $\xb_s$ and receiver account $\xb_r$. 

\subsection{Threat Model and Privacy Objectives}
The primary goal is to provide provable privacy protection to the training data $\Xb_T$ and $\Xb_B$ across the parties throughout the execution of the training algorithm.
Specifically, we must protect the details of bank account features from leaking to the transaction party or other bank parties. We must also protect the transaction features from leaking to the bank parties.

We assume that the parties are ``honest but curious": they comply with the training protocol but might seek to deduce others' data from information exchanged during the algorithm's execution. We assume there is no collusion between parties and that communication is secure, so there are no man-in-the-middle attacks. We also assume that the transaction and account data are not maliciously crafted during pre-processing. We do not consider any privacy concerns at that phase. 

\section{Proposed Algorithm}
\label{sec:Proposed Algorithm}
We now present the steps of the Fed-RD algorithm. Pseudo-code is provided in Algorithm~\ref{vflmpc}.

\begin{algorithm}[h]
    \caption{The Fed-RD algorithm.}
    \label{vflmpc}
    \begin{algorithmic}[1]
    \STATE {\textbf{Initialize:}} $\Thetab^0 = [\thetab_T^0, \thetab_{B}^0, \thetab_{F}^0]$
  \FOR {$t \leftarrow 0, \ldots, Q-1$}
        \STATE Sample minibatch $\B^t$ from $(\Xb_T, \yb)$    
        \FOR {each sample $(\xb_t, \xb_r, \xb_s) \in \B^t$}
        \STATE /* Generate embedding for the transaction, sender account, and receiver account */
        \STATE $\eb_t \gets h_T(\thetab_T^t;\xb_t)$
        \STATE $\eb_s \gets h_B(\thetab_B^t; \xb_s),b, \beta)$
        \STATE $\eb_r \gets h_B(\thetab_B^t; \xb_r),b, \beta)$
        \ENDFOR
        \STATE Embeddings shared with active party via Local RDP \emph{(Approach 1)} or PBM + MPC \emph{(Approach 2)}
        \STATE /* Active party updates fusion model parameters */
         \STATE $\thetab_F^{t+1} \gets \thetab_F^t - \eta \Db(\thetab_F^t)$
        \STATE Active party sends $\nabla_{h} \Loss_{\B^t}$ to transaction party
        \STATE Active party sends relevant entries of  $\nabla_{h} \Loss_{\B_i^t}$ to each bank $i$
        \STATE /* Transaction party updates its model parameters */
        \STATE $\thetab_T^{t+1} \gets \thetab_T^t - \eta^t \Db(\thetab_T^t,\nabla_{h} \Loss_{\B^t})$

        \FOR {each bank $i$ in parallel}
        \STATE /* Bank $i$ computes quantized clipped gradient */
        \STATE $g_i \gets \Db(\thetab_{B_i}^t, \nabla_{h} \Loss_{\B_i^t})$
        \STATE $q_i \gets \textbf{PBM}(Clip(g_i, k), b' \beta')$
        \ENDFOR
        \STATE /* At Bank 1 */
        \STATE $\hat{q}^t \gets \sum_{i=1}^K \tilde{g}_i^t$ via MPC
        \STATE  $\tilde{g}^t \leftarrow \frac{1}{\beta' b'} (\hat{q}^t - \frac{bK}{2})$
        \STATE Bank 1 sends $\tilde{g}^t$ to all other banks.
        \FOR {each bank $i$ in parallel}
        \STATE /* Bank $i$ updates its model */
         \STATE $\thetab_{B_i}^{t+1} \gets \thetab_{B_i}^t - \frac{\eta}{K} \tilde{g}^t $
        \ENDFOR 
\ENDFOR
\end{algorithmic}
\end{algorithm}

First, the model parameters $\thetab_T^0$, $\thetab_B^0$, and $\thetab_F^0$ are initialized (line 1). Note that each bank party in Silo $B$ holds an identical copy of $\thetab_B^0$. 
We denote these copies by $\thetab_{B_1}^0, \ldots \thetab_{B_K}^0$.
 The algorithm runs for $Q$ iterations until a desired convergence criterion is met. 
In each iteration, a minibatch of transaction sample IDs, denoted by $\B^t$, is selected from $(\Xb_T,\yb)$ (line 2). 
For each transaction $i \in \B^t$, the transaction party
generates an embedding $\eb_t^i$. The transaction party contacts the sender and receiver banks and tells them to generate the sender account embedding $\eb_s^i$ and the receiver account embedding $\eb_r^i$ (lines 6-8). The embeddings are then shared with the active party (line 10). Sharing the embeddings directly with the active party could potentially reveal information about the training data. 
 
Fed-RD offers two alternatives to protect the privacy of this information:\\
\emph{Approach 1 (Concatenation)}: In Approach 1, Local RDP is implemented using the Gaussian mechanism. 
Each party adds Gaussian noise with mean 0 and variance $\sigma^2$ to each component of their respective embeddings. 
Each party sends its noisy embedding to the active party, which concatenates them and uses this concatenation  as input to the fusion model.\\
\emph{Approach 2 (Summation)}: Approach 2 implements standard RDP using PBM and MPC. Each party uses Algorithm~\ref{pbm.alg} to quantize every component of its respective embedding. The parties compute the component-wise sum of their quantized embeddings using MPC, and this quantized sum $\hat{h}$ is revealed to the active party. The active party estimates the embedding sum as \(\frac{K}{\beta b} (\hat{h} - \frac{bM}{2})\), as discussed in Section~\ref{sec:PBM}. The active party then utilizes this sum as an input to the fusion model.

The active party uses the fusion model to generate the prediction $\hat{y}_i$ for each transaction $i \in \B^t$. Subsequently, the active party computes the loss and updates the fusion model parameters via a descent step, for example, stochastic gradient descent (SGD) or Adam~\cite{KingBa15}, with learning rate $\eta$ (line 12).
Additionally, the active party computes the partial derivative of the loss function with respect to the transaction embeddings (or embedding sums for Approach 2) and sends this partial derivative $\nabla_h \Loss_{\B^t}$ to the transaction party (line 13). 
The active party computes the partial derivative with respect to the bank account embeddings (or embedding sums for Approach 2) and sends each bank the elements of this partial derivative corresponding to accounts held by that bank, denoted by $\nabla_h \Loss_{\B_i^t}$ (line 14). 

The transaction party uses this partial derivative 
to compute the descent step and updates its parameters (line 15). For example, for Approach 1 with SGD, the party  applies the chain rule to update its model parameters as
\begin{equation}
\thetab_T^{t+1} \gets \thetab_T^t - \eta^t \nabla_{\thetab_T}  h_T(\thetab_T^t;\xb_t)  \nabla_{h} \Loss_{\B^t}. \label{chain.eq}
\end{equation}

Concurrently, each bank $i$ computes its descent step $g_i^t$ using the partial derivative from the active party, similar to (\ref{chain.eq}). The bank then clips the components of $g_i^t$ to bound them within the range $[-k, k]$ and quantizes the components of $g_i^t$ using PBM, with parameters $b'$ and $\beta'$, to generate $q_i^t$. The parties compute the component-wise sum of their quantized descent steps using MPC; this sum $\hat{q}^t$ is revealed to bank 1, which dequantizes it to find the estimated sum $\tilde{g}^t$ (lines 17-24). Bank 1 shares $\tilde{g}^t$ with all other banks, and each bank updates its model parameters with the noisy average descent step as
$\thetab_{B_i}^{t+1} \gets \thetab_{B_i}^t - \frac{\eta}{K} \tilde{g}^t$~(line 28).

The privacy parameters $\sigma^2$, $b$, $b'$, $\beta$, and $\beta'$ can be chosen to achieve different levels of privacy at the cost of injecting different amounts of noise into the training process. We give a formal analysis of the privacy of Fed-RD in the next section, followed by an experimental evaluation of privacy and accuracy in Section~\ref{sec:Experiments}.

\section{Privacy}
\label{privcy}
This section presents a theoretical analysis of Algorithm~\ref{vflmpc}. We first review the potential information leaks and mitigation strategies. We then formally state the privacy guarantees of Fed-RD with Approach 1 and Approach 2, respectively, followed by a discussion of their implications. All proofs are deferred to Appendix~\ref{proof.app}.

\subsection{Information Sharing} There are three potential information leaks in Algorithm~\ref{vflmpc}. The first occurs when the active party learns the embeddings. For Approach 1,  we apply Gaussian noise to each embedding before it is shared to guarantee Local RDP. For Approach 2, we provide DP using  PBM and MPC. The second potential leak occurs when the active party sends $\nabla_{h} \mathcal{L}_{\mathcal{B}^t}$ to other parties during backpropagation. Due to the post-processing property of DP~\cite{dwork2014algorithmic}, the privacy protection of forward propagation is preserved during backpropagation. Finally, a potential leak occurs when banks average their descent steps. We address this privacy concern by using PBM and MPC.

\subsection{Privacy Analysis of Approach 1}
We first give the privacy budget for Fed-RD with Approach~1, where Local RDP is used to share the embeddings with the active party. 
\begin{theorem} \label{approach1.thm}
Let $\beta' \in [0, \frac{1}{4}]$, $b' \in \mathbb{N}$, and  $\alpha \leq 2$. Algorithm~\ref{vflmpc}, after $Q$ iterations, provides $(\alpha, \epsilon_T(\alpha))$-RDP for transactions features with 
\[
\epsilon_T(\alpha) = O\left(\frac{Q P B M_T \alpha}{N\sigma^2}\right)
\]
and provides  $(\alpha, \epsilon_B(\alpha))$-RDP for bank account features with 
\[
\epsilon_B(\alpha) = O\left(\max \left(\frac{Q P B M_T \alpha}{N\sigma^2}, \frac{QB^2}{N^2} \cdot \frac{|\theta_B| b' (\beta')^2 \alpha}{K-1} \right) \right).
\]
where $P$ is the embedding size, $B$ is the transaction batch size, $M_T$ is the maximum number of transactions in which a single bank account participates, and $|\theta_B|$ is the number of parameters in the bank account model.
\end{theorem}

The transaction features and the account features have different privacy budgets due to the different mechanisms for information sharing in Fed-RD. The forward propagation and backward propagation steps utilize the Local RDP Gaussian mechanism, which gives the same privacy budget for both transaction and account feature sets. This is the first argument in the max function for $\epsilon_B(\alpha)$. The second argument is the privacy budget for computing the sum of the bank descent steps, where privacy is achieved via PBM and MPC. Note that both of these privacy budgets can be tuned through the selection of values for $b'$, $\beta'$, and $\sigma^2$.

\subsection{Privacy Analysis of Approach 2}
We now give the privacy budget for the transaction and bank account features for Fed-RD with Approach 2, where embeddings are summed.
\begin{theorem} \label{approach2.thm}
Let $ \beta, \beta' \in [0, \frac{1}{4}]$, $b, b' \in \mathbb{N}$, and  $\alpha \leq 2$. Algorithm~\ref{vflmpc}, after $Q$ iterations, provides $(\alpha, \epsilon_T(\alpha))$-RDP for transactions features with 
\[
\epsilon_T(\alpha) = O\left(\frac{QB}{N} \cdot \frac{P b \beta^2 \alpha}{2}\right)
\]
and provides  $(\alpha, \epsilon_B(\alpha))$-RDP for bank account features with 
\[
\epsilon_B(\alpha) = O\left(\frac{Q B M_T}{N} \cdot \frac{P b \beta^2 \alpha}{2}\right)
\]
for 
\[
N > \frac{ 2 |\theta_B| b' (\beta')^2}{M_T P b \beta^2}
\]
where $P$ is the embedding size, $B$ is the transaction batch size, $M_T$ is the maximum number of transactions in which a single bank account participates, and $|\theta_B|$ is the number of parameters in the bank account model.
\end{theorem}
In Approach 2, Fed-RD uses PBM and MPC as the only mechanism for DP. For forward propagation for a single transaction, the MPC is across three parties: the transaction party, the sender bank, and the receiver bank. When the banks sum their descent steps, this MPC is over $K$ parties, and further, each bank's descent step is an average over $B$ samples. Thus, this sum computation is inherently more private.  When there are a sufficiently large number of training samples, the privacy loss of forward propagation/backpropagation dominates, as shown in the theorem.

\subsection{Discussion}\label{sec:discuss}
According to RDP, a smaller $\epsilon$ implies greater privacy protection. To bolster the privacy assurance of Algorithm~\ref{vflmpc}, one can increase the randomization in the privacy mechanisms by tuning the appropriate parameters. Other factors such as the number of iterations $Q$, batch size $B$, embedding size $P$, and the number of samples $N$ also influence the overall privacy budget of Algorithm~\ref{vflmpc}. However, these parameters are generally set to maximize algorithmic utility and are typically treated as constants in calculating the privacy budget.

Note that because Approach 1 relies on Local RDP, more ``noise'' is required to provide the same privacy protection as Approach 2.
This noise may adversely affect the accuracy of the training. While Approach 2 requires less noise, this comes at the expense of a less flexible fusion model since embeddings are summed before they are input into the fusion model. In the next section, we explore the tradeoffs in privacy and accuracy through experiments.

\section{Experiments}
\label{sec:Experiments}
In this section, we present an experimental evaluation of Fed-RD on realistic synthetic datasets. Additional experimental results are given in Appendix~\ref{sec:addexp}.

\paragraph*{SWIFT dataset} This dataset was provided by SWIFT as part of the NSF-organized Privacy Enhancing Technologies Prize Challenge on Financial Crime Prevention~\cite{nist-challenge}. The training set consists of approximately five million transactions, with a ratio of positive to negative samples maintained at $1:952$. The test set is sized at one-quarter of the training set.
There are 19 transaction features and 3 bank account features, with accounts distributed among 16 banks. 

\paragraph*{AMLSim dataset} This dataset was generated using AMLSim~\cite{altman2024realistic}, a multiagent simulator 
that generates realistic transactions for the investigation of money laundering. The dataset comprises one million transactions, with 5 transaction features and 11 bank account features, with accounts distributed among 13 banks. The ratio of positive to negative samples is 1:295. 

\paragraph*{Comparison with baselines} We compare Fed-RD with several baselines. First, we compare it with an XGBoost model trained only on transaction features. We also compare with a version of Fed-RD that was implemented without any privacy mechanisms.

For the SWIFT dataset, we set the batch size at $B=1000$ and the embedding size at $P=64$.  For the AMLSim dataset, we use a smaller batch size of $B=128$ and use $P=64$. Unless stated otherwise, we use the Adam optimizer with a learning rate of $0.001$.  For both datasets, model accuracy is evaluated using 
the Area Under the Precision-Recall Curve (AUPRC) metric. Details of the neural network architectures are given in Appendix~\ref{app:nn}.

\subsection{Accuracy vs. Privacy}
We first investigate how the privacy mechanisms impact the accuracy of the trained model. We set $b=64$ and $b'=1024$ and explore $\beta$ and $\beta'$ in the set $\{0.10, 0.15, 0.25\}$. For Approach 1 (concatenation), we set $\sigma^2 = 4/(b \beta^2)$ so that Approach 1 and Approach 2 yield the same privacy budget. We run each instance of the training algorithm until it achieves its maximum accuracy.

\begin{table}
\caption{Maximum AUPRC on the SWIFT dataset for various privacy levels.}\label{table:comarison}
\begin{center}
\begin{tabular}{| p{1.9cm} | p{1.6cm} | p{1.6cm} |}
\hline
{\bf Approach} & $\beta$ & {\bf Max AUPRC} \\
\hline
XGBoost & N/A  & 60\% \\
\hline
& No Privacy & 79\%\\
  & 0.10  & 70\%\\
Concatenation  & .15  & 73\%\\
  & 0.25  & 77\%\\
\hline
& No Privacy & 80\%\\
   & 0.10  & 73\%\\
Summation & 0.15  & 75\%\\
   & 0.25 & 78\% \\
\hline
\end{tabular}
\vspace{-5ex}
\end{center}
\end{table}

The results are shown in Table \ref{table:comarison}. We observe that all instances of Fed-RD, both with and without privacy mechanisms, outperform the XGBoost baseline. This demonstrates the benefit of incorporating bank account data into the model for higher accuracy. We also observe that Approach 2 (summation) slightly outperforms Approach 1 (concatenation). As mentioned in Section~\ref{sec:discuss}, Approach 1 requires more noise to provide the same degree of privacy protection, and these results show that this additional noise degrades the model performance. We also note that, as expected, as $\beta$ increases and privacy decreases, the max AUPRC increases. Overall, the performance of privacy-preserving Fed-RD is quite similar to that without privacy, indicating that we can achieve good model performance with strong privacy guarantees.

\begin{figure*}[t] 
\centering
\begin{subfigure}[t]{0.33\textwidth} 
\centering
  \includegraphics[height=3.7cm] {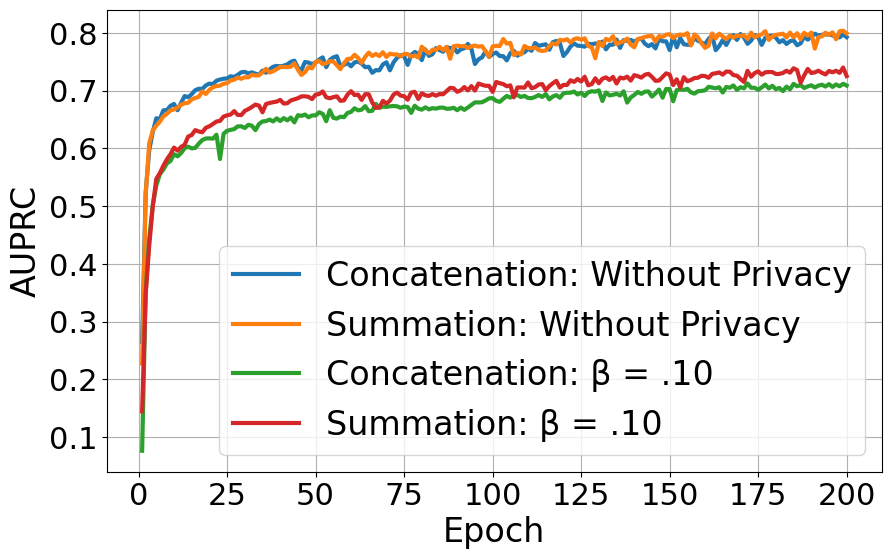} 
  \caption{$\beta = 0.10$}
  \label{subfig1}
\end{subfigure}%
\hfill 
\begin{subfigure}[t]{0.33\textwidth} 
\centering
  \includegraphics[height=3.7cm] {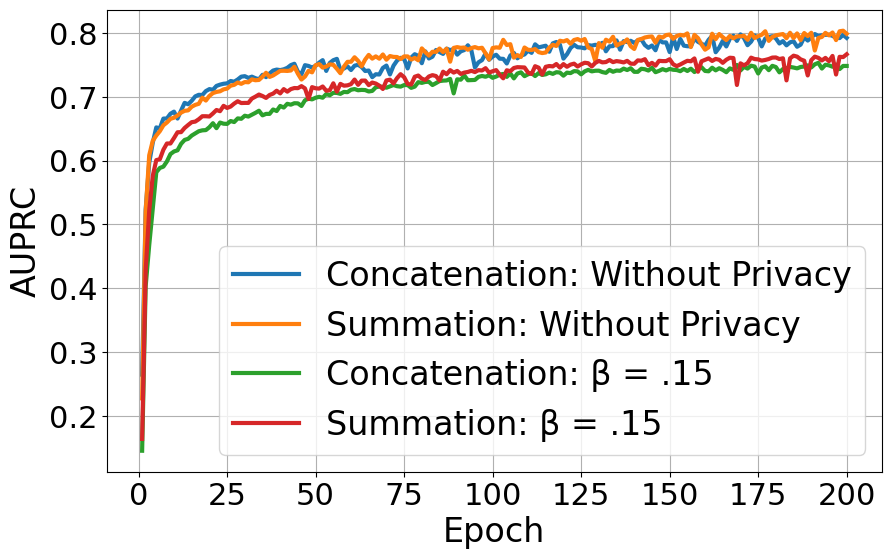} 
  \caption{$\beta = 0.15$}
  \label{subfig2}
\end{subfigure}%
\hfill 
\begin{subfigure}[t]{0.33\textwidth} 
\centering
  \includegraphics[height=3.7cm] {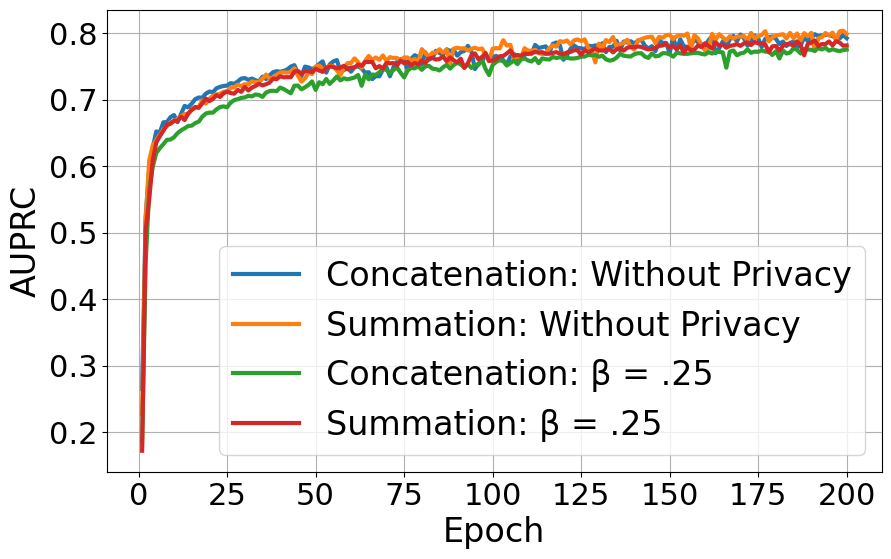} 
  \caption{$\beta = 0.25$}
  \label{subfig3}
\end{subfigure}

\caption{Testing AUPRC on the SWIFT dataset for various values of $\beta$, with $b = 64$ and $b' = 1024$.}
\label{fig:overall}
\end{figure*}

\begin{figure*}[t] 
\centering
\begin{subfigure}[t]{0.33\textwidth} 
\centering
  \includegraphics[height=3.7cm] {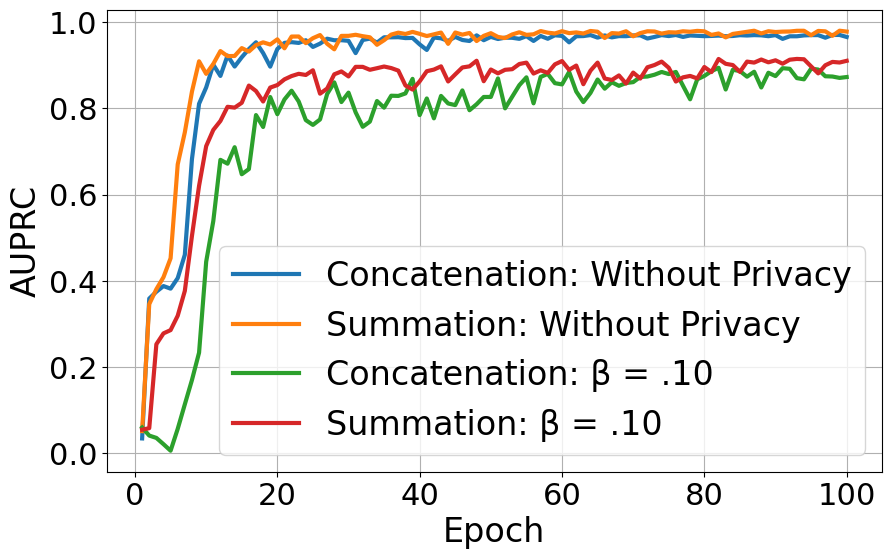} 
  \caption{$\beta = 0.10$}
  \label{aml1}
\end{subfigure}%
\hfill 
\begin{subfigure}[t]{0.33\textwidth} 
\centering
  \includegraphics[height=3.7cm] {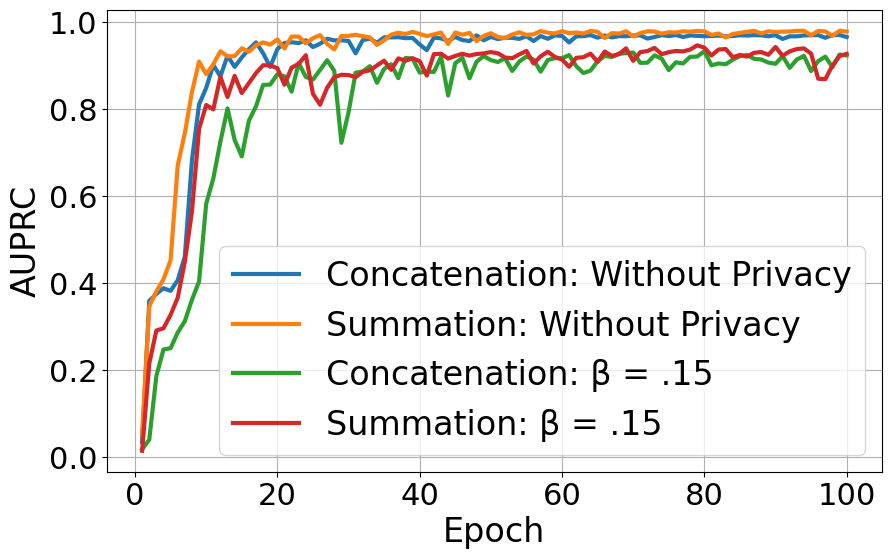} 
  \caption{$\beta = 0.15$}
  \label{aml2}
\end{subfigure}%
\hfill 
\begin{subfigure}[t]{0.33\textwidth} 
\centering
  \includegraphics[height=3.7cm] {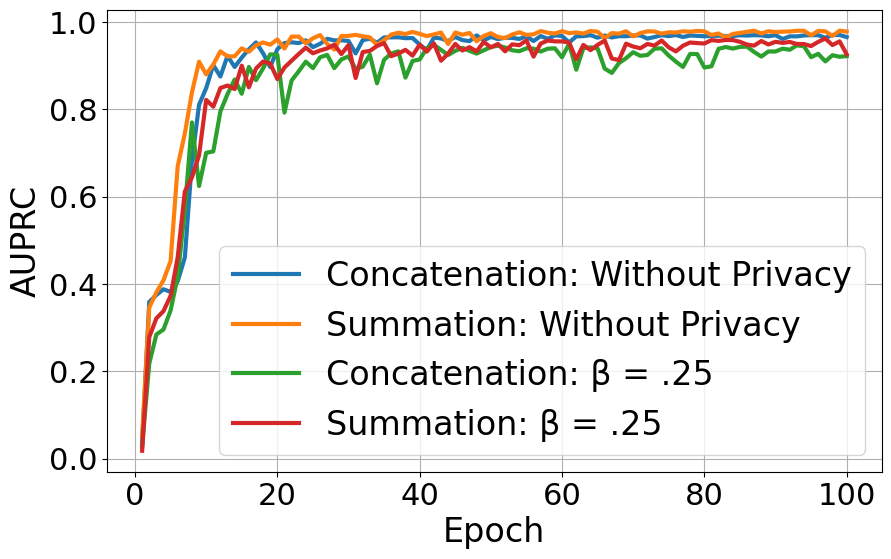} 
  \caption{$\beta = 0.25$}
  \label{aml3}
\end{subfigure}
\caption{Testing AUPRC on the AMLSim dataset for various values of $\beta$, with $b = 64$ and $b' = 1024$.} \label{aml}
\end{figure*}

\subsection{Convergence}
We next study the convergence behavior of Fed-RD for various privacy levels, as determined by the value of $\beta$.
We again set $b=64$ and $b'=1024$ and set $\sigma^2$ to equalize the privacy guarantees of Approach 1 and Approach 2. 

Figure \ref{fig:overall} shows the experimental results on the SWIFT dataset. In Figure~\ref{subfig1}, we present results for a higher privacy guarantee, with $\beta=0.10$. While all versions of the algorithm converge at a similar rate, the AUPRC is lower for the privacy-preserving versions of Fed-RD compared to those that do not provide privacy. 
This difference is because DP adds noise, which impacts the model's accuracy. We further observe that the summation approach slightly outperforms the concatenation approach in this experiment, which is in line with the previous experiments. In Figures~\ref{subfig2} and ~\ref{subfig3}, we compare the same methods but with lower privacy guarantees, $\beta = 0.15$ and $\beta = 0.25$.  As can be seen, as $\beta$ increases, the privacy-preserving Fed-RD performs more similarly to the non-private implementations.
This is as expected because the privacy mechanisms inject less noise.

Figure \ref{aml} shows the test AUPRC score by epoch for the AMLSim dataset.
While the peak AUPRC is higher for this dataset, the overall trends are similar. The Fed-RD implementations without privacy achieve slightly higher AUPRC than those with privacy guarantees, with this performance gap decreasing as $\beta$ increases and privacy decreases.

\begin{table}
\caption{SWIFT dataset - test accuracy target 70\%.} \label{table:perepochtimes}
\centering
\begin{tabular}{| p{1.9cm} | p{0.7cm} | p{1.6cm} | p{2.1cm} |}
\hline
{\bf Approach} & $\beta$ & {\bf Num. Epochs} & {\bf Communication Cost (GB)}\\
\hline\hline
  & 0.10 &  120 & 2229.4\\
Concatenation  & 0.15 & 48 & 891.7\\
  & 0.25 & 27 & 501.6\\
\hline
   & 0.10 & 50 & 166.3\\
Summation & 0.15 & 30 & 99.7\\
   & 0.25 & 19 & 63.1 \\
\hline
\end{tabular}
\end{table}

\begin{table}
\caption{AMLSim dataset - test accuracy target 90\%.}\label{table:aml}
\centering
\begin{tabular}{| p{1.9cm} | p{0.7cm} | p{1.6cm} | p{2.1cm} |}
\hline
{\bf Approach} & $\beta$ & {\bf Num. Epochs} & {\bf Communication Cost (GB)}\\
\hline\hline
  & 0.10 &  - & -\\
Concatenation  & 0.15 & 30 & 20.4\\
  & 0.25 & 16 & 11.1\\
\hline
   & 0.10 & 47 & 5.9\\
Summation & 0.15 & 25 & 3.1\\
   & 0.25 & 15 & 1.9 \\
\hline
\end{tabular}
\vspace{-3ex}
\end{table}

\subsection{Communication Cost}
Finally, we compare the communication costs of Fed-RD for the two approaches and various privacy parameters. Details of how these costs are computed are given in Appendix~\ref{sec:communication}. We note that since Approach 2 (summation) employs quantization during forward propagation, significantly fewer bits are  sent to the active party than in
Approach 1 (concatenation), resulting in a smaller communication cost per epoch.

We summarize the communication costs for Fed-RD to reach a target AUPRC of 70\% on the SWIFT dataset in Table \ref{table:perepochtimes}.  The results illustrate that employing the summation rather than concatenation significantly lowers communication costs. Notably, as $\beta$ increases, a reduction in DP noise facilitates quicker convergence, thus requiring fewer epochs to achieve the target accuracy and further diminishing the communication cost. However, this efficiency comes at the expense of privacy, as a higher $\beta$ translates to lower privacy guarantees.

Table \ref{table:aml} illustrates the communication cost required for Fed-RD to achieve a target AUPRC of 90\% on the AMLSim dataset. It is noteworthy that in the concatenation approach with $\beta = 0.10$, Fed-RD fails to reach the target AUPRC due to high noise levels. Overall, we observe the same trends as for the SWIFT dataset.

\section{Related Work}
\label{relatedwork}
Researchers have widely adopted MPC and DP to protect data or models in FL. \cite{li2023fedvs} and \cite{lu2020multi} propose using MPC protocols for VFL but do not integrate DP during training. Therefore, information leakage is still possible from the results of the MPC computations.  \cite{tran2023privacy} provides a VFL algorithm that guarantees end-to-end privacy through a combination of PBM and MPC.  Privacy protection for training data has also been explored in HFL. 
For example, \cite{agarwal2018cpsgd} combines DP and MPC to protect privacy during the aggregation of party gradients. However, it does not provide end-to-end privacy guarantees.  \cite{chen2023communication} utilizes PBM and MPC  to protect gradient computations and also proves end-to-end data privacy for HFL training. 
It is important to note that although we utilize the same mechanisms as \cite{tran2023privacy} and \cite{chen2023communication}, there are significant differences in their application and privacy analysis. These works consider data distributions that are either horizontal or vertical, but not both simultaneously, as is the case in our setting. Our unique data partitioning necessitates a novel application of PBM and MPC, as well as  new privacy analysis. Recent work proposes TDCD~\cite{das2022cross}, an FL algorithm for settings where data is vertically partitioned across parties and then horizontally partitioned within them. This bears some similarity to the Fed-RD setting. However, TDCD does not protect the privacy of training data. Recent works have proposed various solutions to the problem studied in this work.  For example, \cite{arora2023privacy} develops a hybrid privacy-preserving framework by partitioning data horizontally and vertically and enhancing privacy by employing MPC and DP. The proposed work is limited to the SWIFT dataset, as it relies solely on the `receiver flag' feature within bank datasets, whereas our approach generalizes to different datasets and different account features.
In addition, the work assumes that either the aggregator or the active party are attackers, but not both, which is not a limitation in Fed-RD.
\cite{zhang2023privacy} also employs a separate bank account model but uses an autoencoder rather than supervised training. 
The account model parameters are averaged at the active party, and the average model is shared with the banks. No privacy mechanisms are employed in this step, and information leakage is possible. Similar to Fed-RD, \cite{zhang2023privacy} uses Gaussian noise to protect account embeddings. However, the work does not provide formal privacy analyses or guarantees for this sharing. In contrast, in Fed-RD, we protect all sites from potential information leaks and provide formal privacy guarantees for end-to-end training.

\section{Conclusion}
\label{sec:Conclusion}
Fed-RD presents a robust theoretical foundation for privacy-preserving FL in financial crime detection. We provided formal definitions of data privacy within this setting and gave a theoretical analysis of the privacy of end-to-end training. We then presented experimental results demonstrating the tradeoffs among accuracy, privacy, and communication costs. 

The practical deployment of Fed-RD necessitates addressing stringent regulatory requirements such as General Data Protection Regulation (GDPR), 
California Consumer Privacy Act (CCPA), Bank Security Act (BSA), and Anti-Money Laundering (AML) laws, which mandate compliance while ensuring privacy and security across jurisdictions~\cite{lux2020new}. Fed-RD incorporates advanced privacy-preserving techniques, including DP and MPC, and can handle both vertically and horizontally partitioned data for collaborative analysis without direct data sharing, to meet these regulatory standards. To handle real-time data processing, Fed-RD must be capable of efficiently managing and processing continuous streams of transactional data. This involves optimizing the algorithm to support incremental learning and real-time updates without compromising privacy guarantees.
Another important concern is privacy protection for labels.
We will explore such extensions in future work.

\bibliographystyle{IEEEtran}
\bibliography{references}

\begin{thebibliography}{10}
\providecommand{\url}[1]{#1}
\csname url@samestyle\endcsname
\providecommand{\newblock}{\relax}
\providecommand{\bibinfo}[2]{#2}
\providecommand{\BIBentrySTDinterwordspacing}{\spaceskip=0pt\relax}
\providecommand{\BIBentryALTinterwordstretchfactor}{4}
\providecommand{\BIBentryALTinterwordspacing}{\spaceskip=\fontdimen2\font plus
\BIBentryALTinterwordstretchfactor\fontdimen3\font minus \fontdimen4\font\relax}
\providecommand{\BIBforeignlanguage}[2]{{%
\expandafter\ifx\csname l@#1\endcsname\relax
\typeout{** WARNING: IEEEtran.bst: No hyphenation pattern has been}%
\typeout{** loaded for the language `#1'. Using the pattern for}%
\typeout{** the default language instead.}%
\else
\language=\csname l@#1\endcsname
\fi
#2}}
\providecommand{\BIBdecl}{\relax}
\BIBdecl

\bibitem{4}
{UNODC}, ``{UNO} on drugs and crime. money laundering,'' \url{https://www.unodc.org/unodc/en/money-laundering/overview.html}, accessed: 2024-04-30.

\bibitem{nasdaq2024}
{Nasdaq Verafin}, ``{Nasdaq Verafin} 2024 global financial crime report,'' \url{https://www.nasdaq.com/global-financial-crime-report}, accessed: 2024-04-30.

\bibitem{alsuwailem2023performance}
A.~A.~S. Alsuwailem, E.~Salem, and A.~K.~J. Saudagar, ``Performance of different machine learning algorithms in detecting financial fraud,'' \emph{Computational Economics}, vol.~62, no.~4, pp. 1631--1667, 2023.

\bibitem{yi2023fraud}
Z.~Yi, X.~Cao, X.~Pu, Y.~Wu, Z.~Chen, A.~T. Khan, A.~Francis, and S.~Li, ``Fraud detection in capital markets: A novel machine learning approach,'' \emph{Expert Systems with Applications}, vol. 231, p. 120760, 2023.

\bibitem{fanai2023novel}
H.~Fanai and H.~Abbasimehr, ``A novel combined approach based on deep autoencoder and deep classifiers for credit card fraud detection,'' \emph{Expert Systems with Applications}, vol. 217, p. 119562, 2023.

\bibitem{rahman2023federated}
A.~Rahman, M.~S. Hossain, G.~Muhammad, D.~Kundu, T.~Debnath, M.~Rahman, M.~S.~I. Khan, P.~Tiwari, and S.~S. Band, ``Federated learning-based ai approaches in smart healthcare: concepts, taxonomies, challenges and open issues,'' \emph{Cluster computing}, vol.~26, no.~4, pp. 2271--2311, 2023.

\bibitem{van2023markets}
T.~van~der Linden and T.~Shirazi, ``Markets in crypto-assets regulation: Does it provide legal certainty and increase adoption of crypto-assets?'' \emph{Financial innovation}, vol.~9, no.~1, p.~22, 2023.

\bibitem{mcmahan2017communication}
B.~McMahan, E.~Moore, D.~Ramage, S.~Hampson, and B.~A. y~Arcas, ``Communication-efficient learning of deep networks from decentralized data,'' in \emph{Artificial intelligence and statistics}, 2017, pp. 1273--1282.

\bibitem{luo2021feature}
X.~Luo, Y.~Wu, X.~Xiao, and B.~C. Ooi, ``Feature inference attack on model predictions in vertical federated learning,'' in \emph{Proc. Int. Conf. Data Engineering}, 2021, pp. 181--192.

\bibitem{yazdinejad2023ap2fl}
A.~Yazdinejad, A.~Dehghantanha, and G.~Srivastava, ``Ap2fl: auditable privacy-preserving federated learning framework for electronics in healthcare,'' \emph{IEEE Transactions on Consumer Electronics}, 2023.

\bibitem{sanchez2024federatedtrust}
P.~M.~S. S{\'a}nchez, A.~H. Celdr{\'a}n, N.~Xie, G.~Bovet, G.~M. P{\'e}rez, and B.~Stiller, ``Federatedtrust: A solution for trustworthy federated learning,'' \emph{Future Generation Computer Systems}, vol. 152, pp. 83--98, 2024.

\bibitem{zhang2023multi}
P.~Zhang, N.~Chen, S.~Li, K.-K.~R. Choo, C.~Jiang, and S.~Wu, ``Multi-domain virtual network embedding algorithm based on horizontal federated learning,'' \emph{IEEE Transactions on Information Forensics and Security}, 2023.

\bibitem{castiglia2023flexible}
T.~Castiglia, S.~Wang, and S.~Patterson, ``Flexible vertical federated learning with heterogeneous parties,'' \emph{IEEE Transactions on Neural Networks and Learning Systems}, 2023.

\bibitem{liu2024vertical}
Y.~Liu, Y.~Kang, T.~Zou, Y.~Pu, Y.~He, X.~Ye, Y.~Ouyang, Y.-Q. Zhang, and Q.~Yang, ``Vertical federated learning: Concepts, advances, and challenges,'' \emph{IEEE Transactions on Knowledge and Data Engineering}, 2024.

\bibitem{dwork2014algorithmic}
C.~Dwork, A.~Roth \emph{et~al.}, ``The algorithmic foundations of differential privacy,'' \emph{Foundations and Trends in Theoretical Computer Science}, vol.~9, no. 3--4, pp. 211--407, 2014.

\bibitem{cramer2015secure}
R.~Cramer, I.~B. Damg{\aa}rd \emph{et~al.}, \emph{Secure multiparty computation}.\hskip 1em plus 0.5em minus 0.4em\relax Cambridge University Press, 2015.

\bibitem{alvim2018local}
M.~Alvim, K.~Chatzikokolakis, C.~Palamidessi, and A.~Pazii, ``Local differential privacy on metric spaces: optimizing the trade-off with utility,'' in \emph{Proc. IEEE 31st Computer Security Foundations Symposium}, 2018, pp. 262--267.

\bibitem{mironov2017renyi}
I.~Mironov, ``R{\'e}nyi differential privacy,'' in \emph{Proc. IEEE 30th Computer Security Foundations Symposium}, 2017, pp. 263--275.

\bibitem{chen2022poisson1}
W.-N. Chen, A.~Ozgur, and P.~Kairouz, ``The poisson binomial mechanism for unbiased federated learning with secure aggregation,'' in \emph{Proc. Int. Conf. Machine Learning}, 2022, pp. 3490--3506.

\bibitem{tran2024differentially}
L.~Tran, S.~Chari, M.~S.~I. Khan, A.~Zachariah, S.~Patterson, and O.~Seneviratne, ``A differentially private blockchain-based approach for vertical federated learning,'' \emph{arXiv preprint arXiv:2407.07054}, 2024.

\bibitem{bonawitz2016practical}
K.~Bonawitz, V.~Ivanov, B.~Kreuter, A.~Marcedone, H.~B. McMahan, S.~Patel, D.~Ramage, A.~Segal, and K.~Seth, ``Practical secure aggregation for federated learning on user-held data,'' \emph{arXiv preprint arXiv:1611.04482}, 2016.

\bibitem{KingBa15}
D.~Kingma and J.~Ba, ``Adam: A method for stochastic optimization,'' in \emph{Proc. Int. Conf. Learning Representations}, 2015.

\bibitem{nist-challenge}
NIST, ``Privacy-enhancing technologies ({PETs}) prize challenge: Advancing privacy-preserving federated learning,'' \url{https://www.nist.gov/itl/applied-cybersecurity/privacy-engineering/collaboration-space/challenges}, accessed: 2024-04-30.

\bibitem{altman2024realistic}
E.~Altman, J.~Blanu{\v{s}}a, L.~Von~Niederh{\"a}usern, B.~Egressy, A.~Anghel, and K.~Atasu, ``Realistic synthetic financial transactions for anti-money laundering models,'' \emph{Advances in Neural Information Processing Systems}, vol.~36, 2024.

\bibitem{li2023fedvs}
S.~Li, D.~Yao, and J.~Liu, ``{FedVS}: Straggler-resilient and privacy-preserving vertical federated learning for split models,'' in \emph{Proc. Int. Conf. Machine Learning}, 2023, pp. 20\,296--20\,311.

\bibitem{lu2020multi}
L.~Lu and N.~Ding, ``Multi-party private set intersection in vertical federated learning,'' in \emph{Int. conf. Trust, Security and Privacy in Computing and Communications}, 2020, pp. 707--714.

\bibitem{tran2023privacy}
L.~Tran, T.~Castiglia, S.~Patterson, and A.~Milanova, ``Privacy tradeoffs in vertical federated learning,'' in \emph{Federated Learning Systems Workshop@{MLSys} 2023}, 2023.

\bibitem{agarwal2018cpsgd}
N.~Agarwal, A.~T. Suresh, F.~X.~X. Yu, S.~Kumar, and B.~McMahan, ``{cpSGD}: Communication-efficient and differentially-private distributed {SGD},'' \emph{Advances in Neural Information Processing Systems}, vol.~31, 2018.

\bibitem{chen2023communication}
W.-N. Chen, A.~Ozgur, G.~Cormode, and A.~Bharadwaj, ``The communication cost of security and privacy in federated frequency estimation,'' in \emph{Proc. Int. Conf. Artificial Intelligence and Statistics}, 2023, pp. 4247--4274.

\bibitem{das2022cross}
A.~Das, T.~Castiglia, S.~Wang, and S.~Patterson, ``Cross-silo federated learning for multi-tier networks with vertical and horizontal data partitioning,'' \emph{ACM Transactions on Intelligent Systems and Technology}, vol.~13, no.~6, pp. 1--27, 2022.

\bibitem{arora2023privacy}
S.~Arora, A.~Beams, P.~Chatzigiannis, S.~Meiser, K.~Patel, S.~Raghuraman, P.~Rindal, H.~Shah, Y.~Wang, Y.~Wu \emph{et~al.}, ``Privacy-preserving financial anomaly detection via federated learning \& multi-party computation,'' \emph{arXiv preprint arXiv:2310.04546}, 2023.

\bibitem{zhang2023privacy}
H.~Zhang, J.~Hong, F.~Dong, S.~Drew, L.~Xue, and J.~Zhou, ``A privacy-preserving hybrid federated learning framework for financial crime detection,'' \emph{arXiv preprint arXiv:2302.03654}, 2023.

\bibitem{lux2020new}
M.~Lux and M.~Shackelford, ``The new frontier of consumer protection: financial data privacy and security,'' \emph{Harvard Kennedy School Mossavar-Rahmani Center for Business and Government}, 2020.

\end{thebibliography}



\appendices 

\section{Proofs of Theorems}\label{proof.app}

\subsection{Proof of Theorem~\ref{approach1.thm}}
We first consider the privacy of transaction features.
As shown in~\cite{mironov2017renyi}, in a single iteration, the local RDP for each transaction embedding in the minibatch is $\epsilon(\alpha) = \frac{P \alpha}{\sigma^2}$, and this privacy is conferred to the underlying transaction features as well.
Over the course of $Q$ iterations, a transaction embedding is used $QB/N$ times, which gives a transaction feature privacy budget of $\epsilon_T(\alpha) = O(\frac{Q P B \alpha}{N\sigma^2})$. 

We next consider the privacy of bank account features. For the forward propagation and backpropagation steps of a single iteration, the privacy budget is the same as for the transaction features, i.e., $\epsilon(\alpha) = \frac{P \alpha}{\sigma^2}$. A single account appears at most $QBM_T/N$ times over $Q$ iterations. By the parallel composition theorem~\cite{dwork2014algorithmic}, this yields a privacy budget of $\epsilon_B(\alpha) = O(\frac{Q P B M_T \alpha}{N\sigma^2})$.

 Finally, we consider the privacy when banks sum their descent steps.
According to~\cite{chen2022poisson1}, 
the privacy budget is 
$\epsilon_{avg}(\alpha) = O\left( \frac{QB^2}{N^2} \cdot \frac{|\theta_B| b' (\beta')^2 \alpha }{K-1}\right)$ 
for $Q$ iterations, where $|\theta_B|$ denotes the number of parameters in the bank account model.
This gives the privacy budget for the bank account features as 
\[
\epsilon_B(\alpha) = O\left(\max \left(\frac{Q P B M_T \alpha}{N\sigma^2}, \frac{QB^2}{N^2} \cdot \frac{|\theta_B| b' (\beta')^2 \alpha }{K-1} \right) \right).
\]

\subsection{Proof of Theorem~\ref{approach2.thm}}
We first consider the privacy of transaction features.
According to~\cite{chen2022poisson1}, for a single transaction in a single iteration, the privacy budget is $\epsilon_T(\alpha) = O\left(\frac{P b \beta^2 \alpha}{2}\right)$.
 In each iteration, a transaction is chosen for training at a rate of $B/N$. Thus, over $Q$ iterations, a transaction appears in training at most $QB/N$ times, leading to a privacy budget of $\epsilon_T(\alpha)~=~O\left(\frac{QB}{N} \cdot \frac{P b \beta^2 \alpha}{2}\right)$ for each transaction. 

 Next, we consider the privacy of the account features in the forward propagation and backward propagation steps.
A single account appears at most $QBM_T/N$ times over $Q$ iterations. This yields a privacy budget of $\epsilon_B(\alpha) = O\left(\frac{Q B M_T}{N} \cdot \frac{P b \beta^2 \alpha}{2}\right)$ for each account.

 Finally, we consider the privacy when banks sum their descent steps. As in the proof of Theorem~\ref{approach1.thm}, the privacy budget is 
$\epsilon_{avg}(\alpha) = O\left( \frac{QB^2}{N^2} \cdot \frac{|\theta_B| b' (\beta')^2 \alpha}{K-1}\right)$ 
for $Q$ iterations, where $|\theta_B|$ denotes the number of parameters in the bank account model.

Assuming $N > \frac{ 2|\theta_B| b' (\beta')^2}{M_T P b \beta^2}$, we have
\[
\frac{M_T P b \beta^2}{2}  > \frac{|\theta_B| b' (\beta')^2}{N(K-1)} .
\]
Thus, 
the privacy budget of the forward propagation and backward propagation steps dominate that of the gradient averaging, yielding the theorem.

\section{Additional Experimental Details}\label{expsetup.app}

\subsection{Neural Network Model Architectures}\label{app:nn}
We use the same neural network model for both datasets.
For the transaction model and bank model, we employ a three-layer neural network with two fully connected layers with 128 and 64 units, respectively.
A normalization follows each fully connected layer. We apply a ReLU activation function after the normalization of the second layer and a $tanh$ activation function after the third.

The fusion model combines the outputs of the transaction model and bank model with an input layer dimension of 192 for Approach 1 and 64 for Approach 2. It comprises a fully connected layer with 32 units, followed by a normalization layer and a sigmoid activation function.
We use cross-entropy loss for all experiments with Fed-RD.

\subsection{Communication Cost} \label{sec:communication}
Let $F$ be the number of bits needed to represent a floating-point number. For the concatenation approach, each party sends a single embedding to the active party at a cost of $PF$ bits. The cumulative cost for 3 parties and batch size $B$ becomes $3BPF$. During backpropagation, the active party transmits the partial derivatives to each party at the same cost of $3BPF$ bits. Therefore, the total communication cost for forward and backpropagation for the concatenation approach is $6QBPF$. 

In the summation approach, utilizing Protocol 0 for Secure Aggregation as detailed in \cite{bonawitz2016practical} for MPC, each party sends its masked quantized embeddings at a cost of $P(\log 3 + \log b)$ bits. For batch size $B$, it becomes $3BP(\log 3 + \log b)$ bits. During backpropagation, the active party transmits the partial derivatives to each party without quantizing them, representing the most costly step in the message exchange. The communication overhead for backpropagation is quantified as $3BPF$ bits. Therefore, the total communication cost for forward and backpropagation for the summation approach is $3QBP(\log 3 + \log b + F)$. 

To sum the bank descent steps, again using Protocol 0, each bank sends its masked quantized values to a corresponding bank at a cost of $|\thetab_{B}|(\log b' + \log K)$ bits. Therefore, the total communication cost for $Q$ iterations with
$K$ banks is $K  | \thetab_B | (\log b' + \log K)$ bits. Subsequently, bank  1 computes and shares the dequantized sum with the other banks, leading to an additional communication cost of $QF(K-1)| \thetab_B |$ bits. 
Thus, the total cost of averaging the gradients is 
$Q|\thetab_B|(K\log b' + K\log K+ (K-1)F)$ bits, and this applies to both approaches.

\begin{figure}[h]
\centering

\includegraphics[scale=0.25]{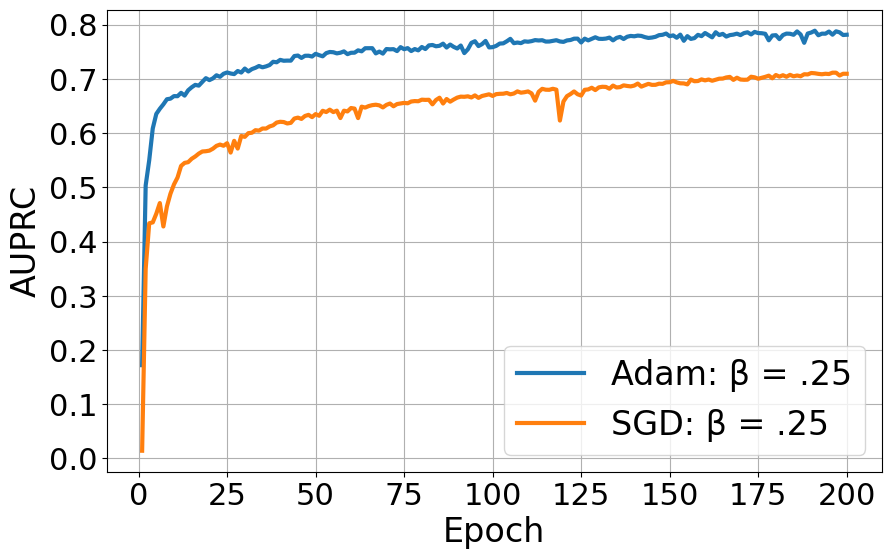}
\caption{Comparison between Adam and SGD optimizer.}
\label{comparison.fig}
\vspace{-3ex}
\end{figure}
\subsection{Additional Experimental Results}\label{sec:addexp}

We further evaluate our Fed-RD using the SGD optimizer and compare its performance with that of the Adam optimizer. We run this experiment for Approach 2. Both optimizers are configured with $\beta$ set to 0.25.   The results are shown in Figure \ref{comparison.fig}. Adam demonstrates faster convergence, reaching higher AUPRC values more quickly than SGD. This rapid progression can be attributed to Adam's momentum components, which help in navigating the parameter space more effectively. In contrast, SGD shows a slower initial increase in performance but demonstrates gradual and consistent improvement. However, it does not surpass the performance of Adam within the span of 200 training epochs.

\end{document}